\documentstyle[aps,prl,preprint]{revtex}
\begin{document}
\draft
\preprint{BARI-TH 347/99}
\date{July 1999}
\title{
PHASE ORDERING IN CHAOTIC MAP LATTICES WITH CONSERVED DYNAMICS}
\author{
Leonardo Angelini, Mario Pellicoro, and Sebastiano Stramaglia}
\address{
Dipartimento Interateneo di Fisica\\
Istituto Nazionale di Fisica Nucleare, Sezione di Bari\\
via Amendola 173, 70126 Bari, Italy}
\maketitle
\begin{abstract}
Dynamical scaling in a two-dimensional lattice model of chaotic maps, in contact with a thermal bath at temperature $T$, is numerically studied. The model here proposed is equivalent to a conserved Ising model with couplings which fluctuate over the same time scale as spin moves.
When couplings fluctuations and thermal fluctuations are both important, this model does not belong to the class of universality of a Langevin equation known as model B; the  scaling exponents are continuously varying with $T$ and depend on the map used. The universal behavior of model B is recovered when thermal fluctuations are dominant. 
\end{abstract}

The kinetics of phase separation as a system is quenched below its ordering temperature is the subject of much current research \cite{datt}. In the dynamical scaling regime the ordering process is characterized by a single time dependent length $R(t)\sim t^{z}$ representing the average domain size \cite{bray}; the growth exponent $z$ usually does not depend on the dimensionality of the system nor on the final temperature of the quenching. In most models with non-conserved order parameter $z=1/2$, while in scalar models with conserved order parameters $z=1/3$ \cite{bray}. A new exponent $\theta$ has been recently introduced by Derrida et al. \cite{der} with the hope to better characterize the process of phase separation: it is related to the persistence probability $p(t)$ defined as the probability that the local order parameter at a given point has never changed since the initial time; tipically one finds $p(t)\sim t^{-\theta}$ \cite{newman}.

In a recent paper phase separation mechanisms have been investigated in the framework of Coupled Chaotic Lattices \cite{lem1}, reaction-diffusion systems in which chaotic maps are coupled diffusively \cite{chate}. In the strong coupling regime these models exhibit nontrivial collective behavior, i.e. long-range order emerging out of local chaos. In \cite{lem1} the one-body probability distribution functions (pdfs) of local (continuous) variables had two disjoint supports; this allowed the unambiguous definition of Ising spin variables. Starting from uncorrelated initial conditions, for large coupling values, complete phase ordering was observed; due to the deterministic nature of the system, the corresponding Ising spin system was at zero temperature. Both the domain size $R(t)$ and the persistence $p(t)$ showed scaling behavior and the exponents $z$ and $\theta$ were found to vary continuously with parameters, at odds with traditional models. {\it Normal} phase ordering behavior, corresponding to the time dependent Ginzburg-Landau equation, was recovered in a continuous space limit of the model \cite {lem2}.
These findings suggest that the phase ordering properties of multiphase Coupled Chaotic Lattices are different from those of most models studied traditionally. It is interesting, therefore, to investigate the nonuniversality of scaling exponents in {\it other} systems of chaotic maps.

In this work, we numerically study the dynamical scaling in a lattice model of chaotic maps such that the corresponding Ising spin model conserves the order parameter. By assuming the system to be in contact with a thermal bath, we also study the competition between the autonomous fluctuations of couplings and thermal fluctuations.

On each site $i$ of a two dimensional square lattice, consider a dynamical system described by the variables $x_i$ and $y_i$, which evolve according to the following map:
\begin{equation}
x_i(t+1)=\mbox{erf} \left({\mu_i (t)\over \sigma_i(t)}\right) - {1\over 2}
\left [\mbox{erf} \left({\tau+\mu_i (t)\over \sigma_i(t)}\right) - 
\mbox{erf} \left({\tau-\mu_i (t)\over \sigma_i (t)}\right) \right ],
\label{eq:flusso1}
\end{equation}

\begin{equation}
y_i(t+1)={1\over 2}
\left [\mbox{erf} \left({\tau+\mu_i (t)\over \sigma_i (t)}\right) + 
\mbox{erf} \left({\tau-\mu_i (t)\over \sigma_i (t)}\right)\right],
\label{eq:flusso2}
\end{equation}

\noindent
where $\mu_i(t)=KJ_0 x_i(t) $, and $\sigma_i (t)=\sqrt{K\left(y_i(t) - J_0^2 x_i^2(t)\right)}$; $K$, $J_0$, and $\tau$ are free parameters. This map has been studied in \cite{caro} in the frame of chaotic neural networks; for a suitable choice of the parameters, it has two symmetric chaotic attractors, one with $x>0$ and the other with $x<0$. Hence an Ising spin $\sigma_i(t) = sign[x_i(t)]$ can be associated to each dynamical system. At each discrete timestep $t$ we assume the following dynamics, consisting of two stages: firstly all the maps are iterated. Secondly, all pairs of nearest neighboring maps are sequentially swapped with the following exchange probability \cite{kawa}: $P_{swap}=1/(1+\exp \{\beta\Delta E_t\})$, where $\beta$ is the inverse temperature, $\Delta E_t$ is the change in energy that would occur if the maps were exchanged, and the configurational energy is defined as $E_t =-\sum_{\langle ij \rangle} x_i(t) x_j(t)$, the sum being over all the nearest neighboring pairs.
It is clear that this dynamics conserves the spin magnetization $\sum \sigma_i$; by writing $E_t =-\sum_{\langle ij \rangle} J_{ij}(t)\sigma_i(t)\sigma_j(t)$, with $J_{ij}(t)=|x_i(t) x_j(t)|$, we observe that this model is equivalent to a conserved Ising model with couplings which fluctuate over the same time scale as spin moves. To study the phase ordering process, uncorrelated initial conditions were generated as follows: one half of the sites were chosen at random and the corresponding values of $x$ and $y$ were assigned according to the invariant distribution of the chaotic attractor with $x>0$, while the other sites were similarly assigned values with $x<0$. Large lattices (up to $1000\times 1000$ sites) with periodic boundary conditions were used; the persistence $p(t)$ and the domain size $R(t)$ were measured. $R(t)$ was estimated by the condition $C(R(t),t)=1/2$, where $C({\bf r},t)=\langle \sigma_{i+{\bf r}} (t)\sigma_i (t) \rangle$ is the two point correlation function. The persistence $p(t)$ was measured as the proportion of sites which have not changed $\sigma$ from the initial time. Both $R(t)$ and $p(t)$ were averaged over many different samples of initial conditions.

Let us now discuss our results. In Fig. 1 typical configurations of the ordering system, at various times, are shown: the morphology of growing domains is similar to those in the conserved Ising model \cite{marko}. At zero temperature ($\beta = \infty$) the system coarsens and exhibits scaling behavior. Choosing $K=10$, $\tau =5$ and $J_0 =0.9$, we measure $z=0.11$ and $\theta =0.59$. In Fig. 2a the time evolution of $R(t)$ is shown, while in Fig. 2b the time evolution of the persistence $p(t)$ is shown. We remark that the conserved Ising model with spin-exchange dynamics (which is recovered by suppressing couplings fluctuations in the present model) does not coarsen at zero temperature \cite{marko}. 
In the case of finite temperature, we find that the growth exponent varies continuously with $\beta$. In Fig. 3 we plot our measures of $z$ for some values of the temperature $T=1/\beta$. We observe that $z$ increases monotonically with $T$ and reaches the value $1/3$ at $T\approx 0.125$; it remains constant until $T=0.32$. Therefore, in the range $T\in [0.125,0.32]$, thermal fluctuations dominate over couplings fluctuations and the system belongs to the class of universality of  a Langevin equation known as model B \cite{gun}, which describes the standard conserved Ising model (when bulk diffusion dominate over surface diffusion \cite{marko}). 
At high temperature ($T > 0.375$) the system does not order and $R(t)$ tends to a constant after a little transient ($z =0$). For temperatures in the range $[0.32,0.375]$, our numerical plots of $R(t)$ do not show a neat scaling behavior (filled area in Fig. 3): probably, in the thermodynamic limit a critical value $\beta_c$ exists, in this range, separating the $z=1/3$ behavior from the $z=0$ one. Further numerical simulations are needed to study the phase ordering properties of this system close to $\beta_c$. We note that the order of magnitude of $\beta_c$ here estimated is consistent with the critical coupling of the two-dimensional Ising model ($\beta_c =0.44$) if one approximates the average value of couplings as follows: $\langle J_{ij}\rangle\approx \langle|x_i(t)|\rangle\langle |x_j(t)|\rangle =0.20$, the average of $x's$ being calculated over the invariant distribution of the chaotic attractor. 

It is worth stressing that simulations of the spin-exchange Ising model, at low temperature, do not provide $z=1/3$ but a value in the range $0.17-0.25$, because of corrections due to excess transport in interfaces \cite{marko}. We cannot exclude that these effects, peculiar to spin-exchange dynamics, still play a role, at low temperature, in the present model. However, we simulated, at low temperatures, the spin-exchange Ising model and, fitting the domain size with the power law form, we measured growth exponents differing from those measured in our model (using the same ratio $T/T_c$ in the two models).

Now we consider the scaling function. In the scaling regime the  correlation function obeys the following scaling form $C(r,t)=f(r/R(t))$.
We remark that recently the universality of the scaling function $f$, with respect to details of the system, has been questioned \cite{rute}.
In Fig. 4 the scaling collapse of the correlation function is depicted for three values of the temperature. At small $r/R$ the scaling function has a linear behavior (Porod's law) $f\approx 1-\alpha \;r/R$; we measure $\alpha \approx 0.5$, independently of the temperature. On the other hand, at larger distances $f$ shows some dependence on the temperature (see Fig.4): 
at the moment, we have no argument to explain this dependence. 
Concerning the persistence exponent, at finite temperature we measure the block persistence $p_{block}(t)$, which, for a sufficiently large block, is much less affected by thermal corrections than $p(t)$  \cite{cuei}. We find that also $\theta$ depends on the temperature; for example at $\beta=20$, using $8\times 8$ blocks, we measure $\theta =1.05$. 

We also simulated systems with different maps. We do not report here details and only quote the results.
We find that, at fixed temperature, the scaling exponents depend on the map. Changing the parameters in Eqs. (1-2) and fixing $K=10$, $\tau =5$, $J_0 =0.95$, we measure, at zero temperature, $z=0.22$ and $\theta =1.57$ ; using the one-dimensional map of Ref. \cite{lem1} (eq. 2), at zero temperature, we find $z=0.07$ and $\theta =0.09$. In both these two cases, however, we verify that the $z$ exponent grows continuously with temperature and, when thermal fluctuations are dominant, $z$ is close to $1/3$. The scaling function $f$ is found to be independent of the map at zero temperature.

In summary, we have numerically studied the dynamical scaling in a lattice model of chaotic maps which is equivalent to a conserved Ising model with fluctuating couplings, in contact with a thermal bath at temperature $T$.
When couplings fluctuations and thermal fluctuations are both important, this model does not belong to the class of universality of model B: the  scaling exponents are continuously varying with $T$ and depend on the map used. The universal behavior of model B is recovered when thermal fluctuations are dominant. Our measures of the scaling function show some dependence on the temperature.  
A similar variation of the scaling exponents with parameters was found in \cite{lem1}, where the Ising models associated to the Coupled Map Lattices were not conserving the order-parameter and at zero temperature: here we have confirmed this behavior in another class of Chaotic Map Lattices, whose corresponding Ising models are conserving the order parameter and in presence of thermal fluctuations.

Finally, we observe that, on the long term, local chaos may be seen as bounded {\it effective} noise. It follows that a similar behavior might be found in the dynamical scaling of conserved coarsening systems in presence of multiplicative noise \cite{garcia}, i.e. external fluctuations of a control parameter in the system. Further investigation, especially at the analytic level, is needed to clarify the origin of the nonuniversality here described.

\section*{Ackowledgements}

We thank G. Gonnella for valuable suggestions.
We also thank D. Caroppo and G. Nardulli for useful discussions.

\newpage
\noindent\Large\textbf{Figure Captions}
\normalsize
\vspace{1.0cm}
\begin{description}
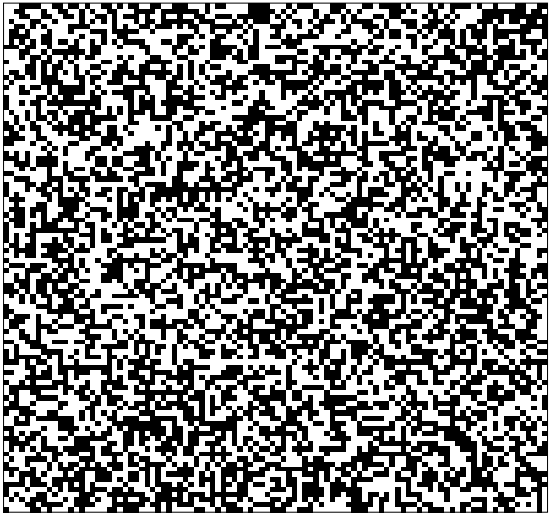
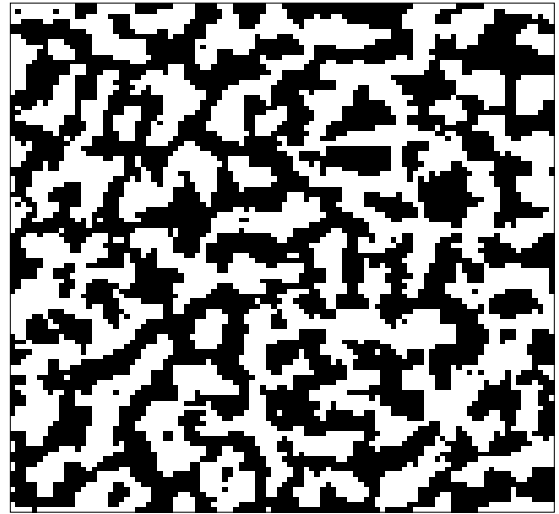
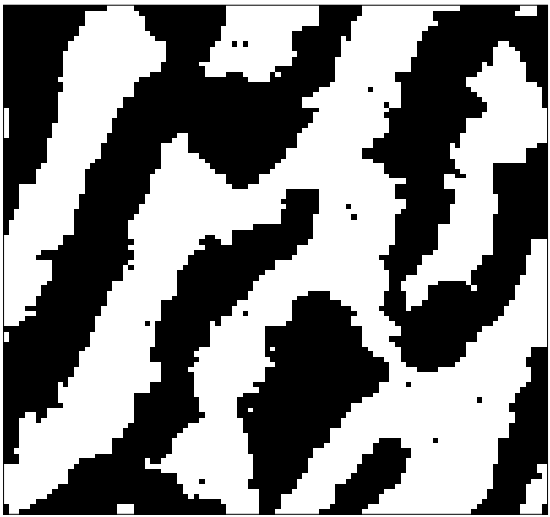
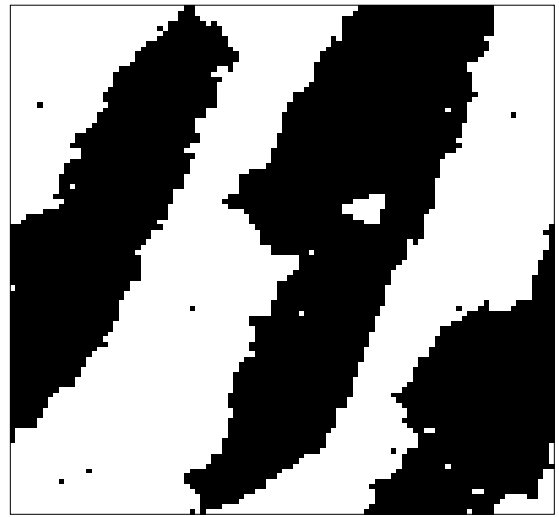
\item{Figure 1}: Snapshots of the ordering system. Black (white) pixels correspond to $\sigma =1$ ($-1$). Lattice of $100\times 100$ sites, $\beta =10$ and iterations times $t=0$ (a), $t=200$ (b), $t=25,000$ (c), $t=1,000,000$ (d).

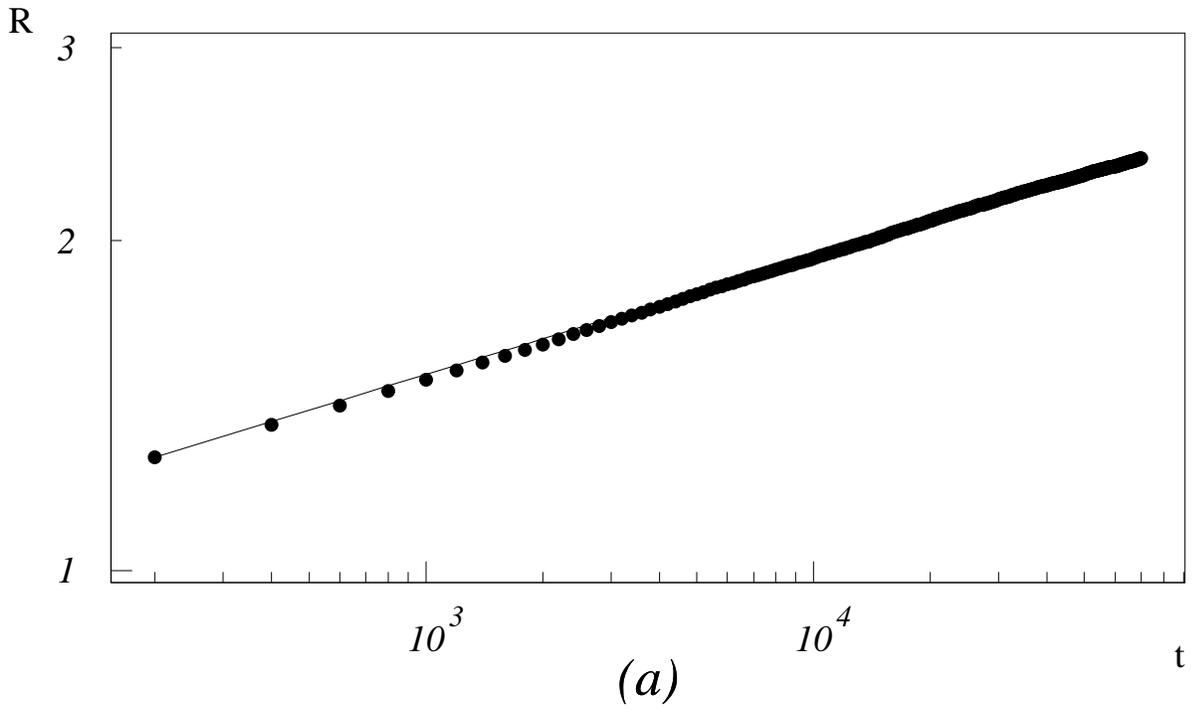
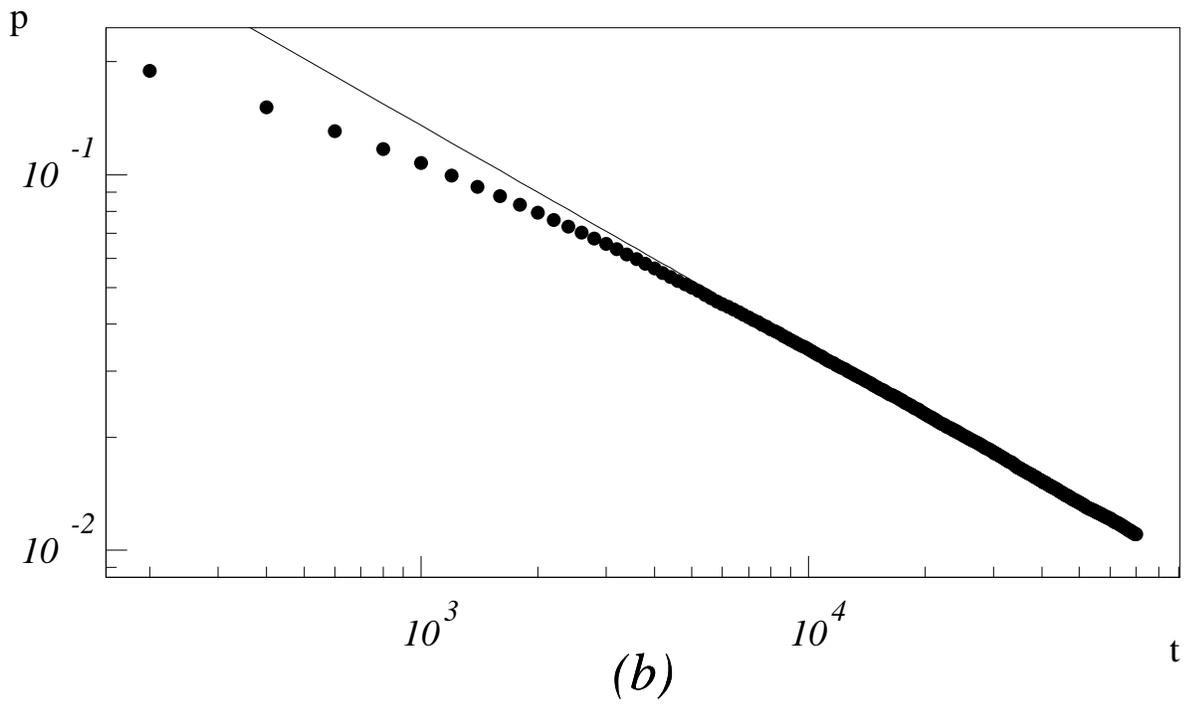
\item{Figure 2}: Time evolution of the domain size $R(t)$ (a) and persistence $p(t)$ (b) at zero temperature. Solid lines are best linear fits. 

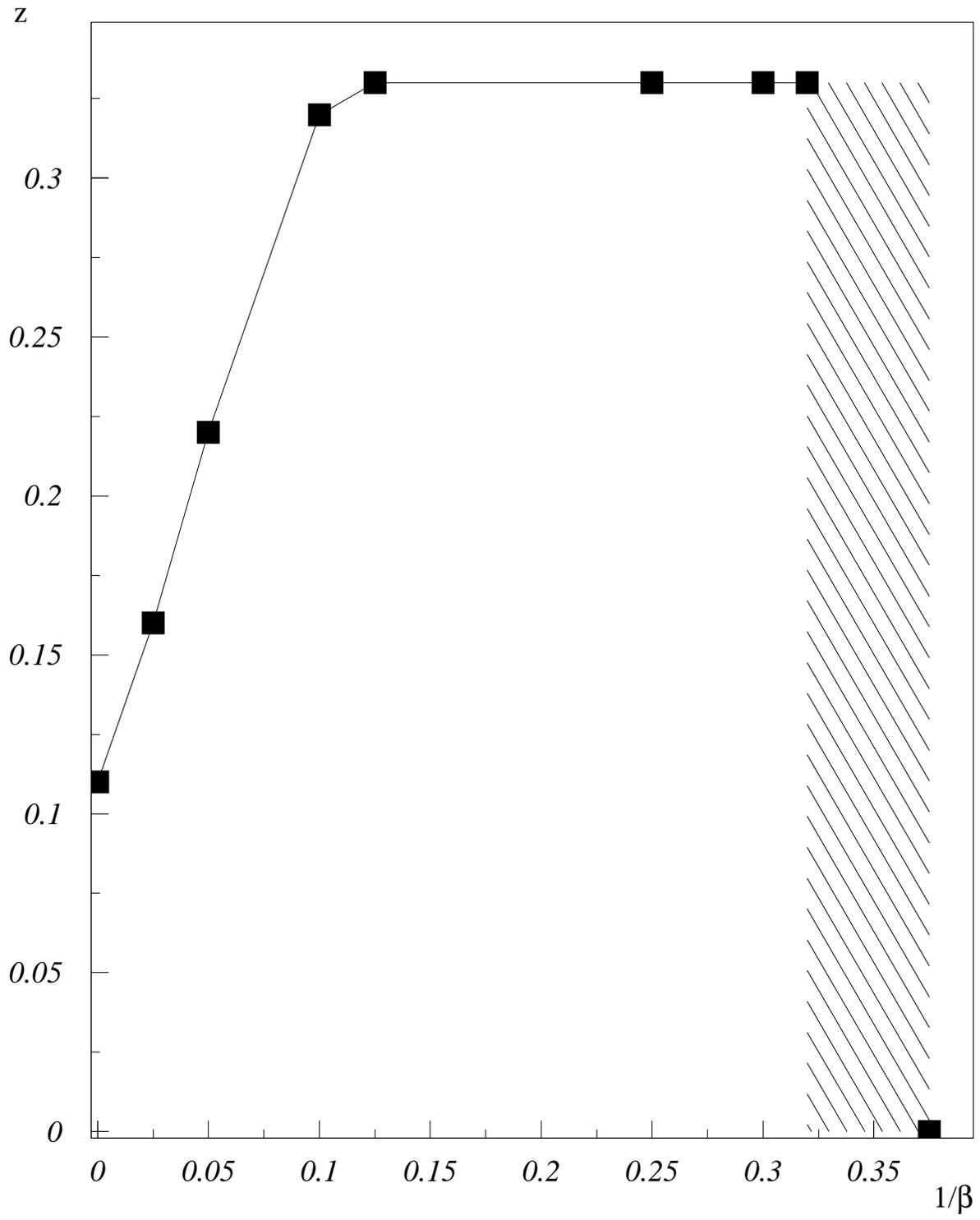
\item{Figure 3}: The estimated growth exponent $z$ versus the temperature $T=1/\beta$.

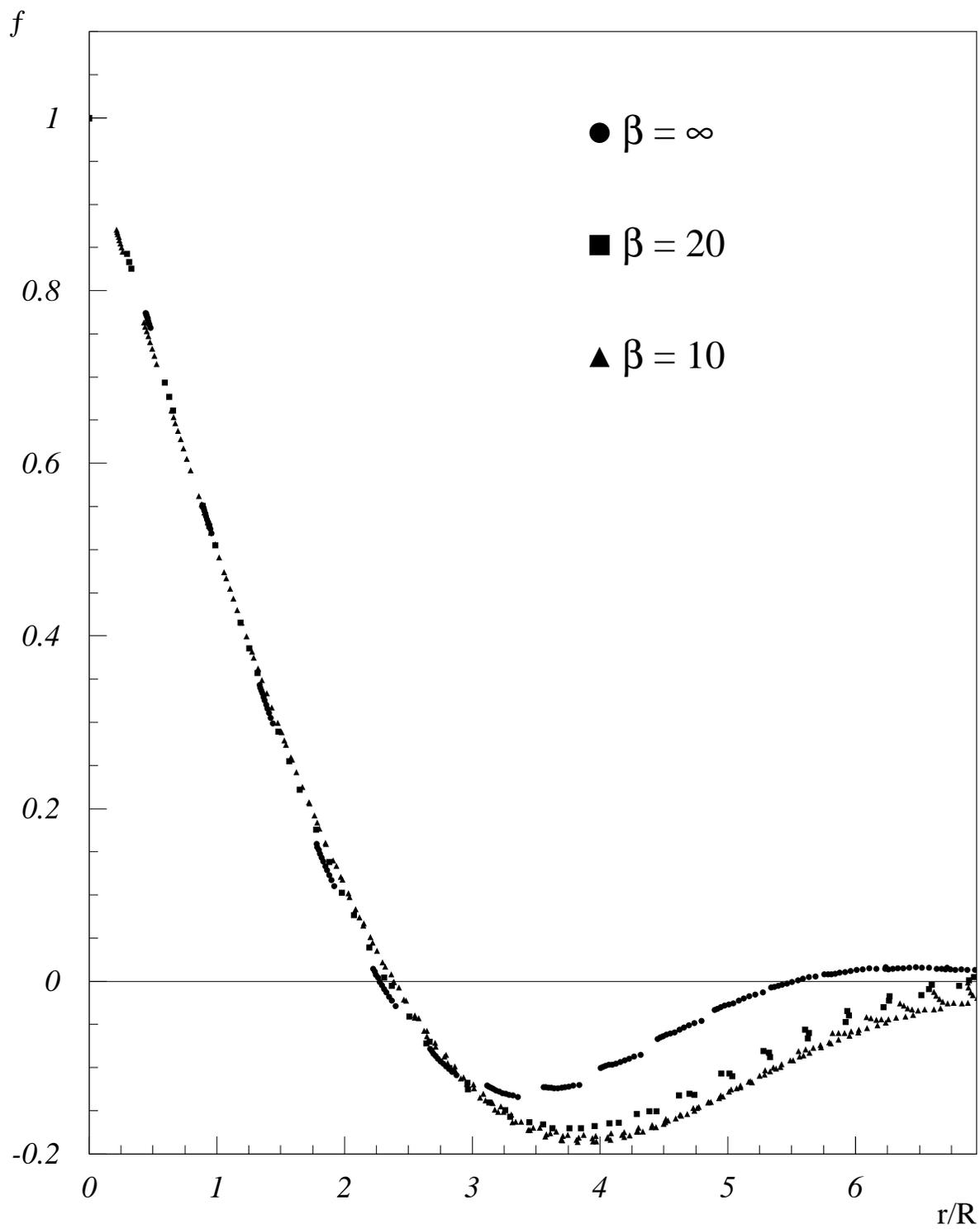
\item{Figure 4}: Scaling collapse of the correlation function for $\beta =\infty$ (circles), $\beta = 20$ (squares) and $\beta= 10$ (triangles). 
Correlations at eight times, equally spaced in the interval [$10^4$, $4\times 10^4$], are shown.

\end{description}
\end{document}